\documentclass[conference]{IEEEtran}
\usepackage{amsmath}
\usepackage{lipsum}
\makeatletter
\newcommand*{\balancecolsandclearpage}{%
  \close@column@grid
  \cleardoublepage
  \twocolumngrid
}
\def\BibTeX{{\rm B\kern-.05em{\sc i\kern-.025em b}\kern-.08em
    T\kern-.1667em\lower.7ex\hbox{E}\kern-.125emX}}

\usepackage{mathrsfs}
\IEEEoverridecommandlockouts
\usepackage{graphicx}
\usepackage{booktabs} 
\usepackage{amsmath,bm}
\usepackage{amsmath}
\usepackage[super]{nth}
	\usepackage[T1]{fontenc}
\usepackage{cite}
\ifCLASSINFOpdf
  \usepackage{graphicx}

    \usepackage{diagbox}

\newcommand*{\rom}[1]{\expandafter\@slowromancap\romannumeral #1@}
\makeatother
\fi
\usepackage{flushend}
\usepackage{multirow}
\usepackage{array}
\usepackage[lofdepth,lotdepth]{subfig}
\usepackage[none]{hyphenat}
\usepackage{xcolor}
\usepackage{amsmath}
\usepackage{amsmath}
\usepackage{amssymb}

\AtBeginDocument{%
  
}

\AtBeginDocument{%
  
}

\begin{document}
\newcolumntype{s}{>{\columncolor[HTML]{AAACED}} p{3cm}}

%
\title{CNN-aided Channel and Carrier Frequency Offset Estimation for HAPS-LEO Links}

\author{
}

\author{\IEEEauthorblockN{Eray Güven, Güneş Karabulut Kurt} 
\IEEEauthorblockA{Poly-Grames Research Center, Department of Electrical Engineering\\ Polytechnique Montr\'eal, Montr\'eal, Canada \\
E-mail : guvenera@itu.edu.tr, gunes.kurt@polymtl.ca
}
}
\maketitle

\begin{abstract}
Low Earth orbit (LEO) satellite mega-constellation networks aim to address the high connectivity demands with a projected 50,000 satellites in less than a decade. To fully utilize such a large-scale dynamic network, an air network composed of stratospheric nodes, specifically high altitude platform station (HAPS), can help significantly with a number of aspects including mobility management. HAPS-LEO network will be subject to time-varying conditions, and in this paper, we introduce an artificial intelligence (AI)-based approach for the unique channel estimation and synchronization problems. First, channel equalization and carrier frequency offset with residual Doppler effects are minimized by using the proposed convolutional neural networks based estimator. Then, the data rate is compounded by increasing spectral efficiency using non-orthogonal multiple access method. We observed that the proposed AI-empowered HAPS-LEO network provides not only a high data throughput per second but also higher service quality thanks to the agile signal reconstruction process. 

\end{abstract}


\IEEEpeerreviewmaketitle

\IEEEpubidadjcol

\section{Introduction}
Air/space networks are becoming  a part of alternative solutions to non-terrestrial networks (NTN) of next generation wireless systems (6G). Satellites are   common NTN elements  that benefit from line of sight (LOS) propagation with a wide coverage. Specifically, low Earth orbit (LEO) constellations, which have lower latency and can work cooperatively and share the workload are experiencing a rejuvenated interest. Cost barrier is decreasing day by day, it is expected that many constellations such as Starlink and Lightwave will become fully operational. \cite{daehnick2020large} estimates that nearly 50,000 active satellites will be placed within a decade. 

Even though mega-constellations are expected to bring solid advantages to aerial networks, they also lead to mobility management and congestion issues by handoffs \cite{CHEN2019376}. As mega-constellation elements need to be seamlessly connected to terrestrial networks, a support infrastructure is required to relieve the load in these networks.  \cite{kurt2021vision} offers a flexible and scalable solution by making use of high altitude platform station (HAPS), stratospheric platforms that are located around 20km altitude. A HAPS provides an efficient and promising solution to handoff load \cite{wang2019effect} in the low complexity nodes (e.g. drone mounted unmanned aerial vehicles (UAVs) and mobile users by taking care of computational weight\cite{alzenad2018fso}). While a study \cite{hsieh2020uav} shows promising performance results for both uplink (UL) and downlink (DL) in aerial networks, $\nth{3}$ Generation Partnership Project (3GPP) Release 17 highlights HAPS based networks \cite{3gpp.36.331} in the sense of NTN. For the recent studies regarding HAPS, the new model was added as a part of a revision to the International Telecommunication Union Radiocommunication Sector Recommendation ITU-R P.1409-2 \cite{itu371.1996}. 

\begin{figure}[t] 
    \centering
    \includegraphics[width=0.45\textwidth]{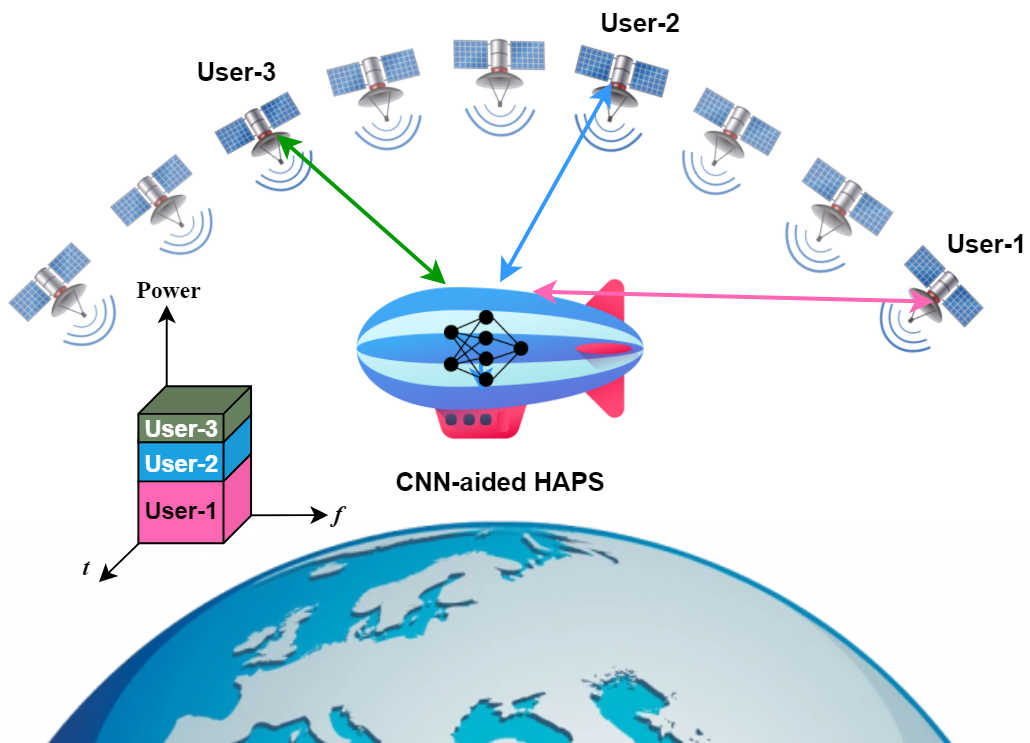}
    \caption{A CNN-aided HAPS-LEO NOMA network.}
    \label{nom}
\end{figure}

In this paper, we focus on HAPS-LEO network's main technical problems (Fig. \ref{nom}),  varying channel gains and the residual Doppler related carrier frequency offset (CFO) effect. Observing that the use of a neural network with an error-minimizing architecture for the solution of non-convex problems is quite functional where unpredictability is implacable \cite{lecun1999object}, we propose a convolutional neural networks (CNN) based approach to detect these parameters as their accurate estimation will be essential for the required performance of  HAPS systems that bridge the satellite and terrestrial networks. The contributions of this study to the literature can be summarized as follows:
\begin{enumerate}
    \item The usability of a HAPS supported satellite constellation in solving mobility and overloading problems is proposed and the flexibility it provides is investigated.
    \item Frequency selective fading channels and non-synchronous carrier frequency reasoned degradation are analyzed for both uplink and downlink.
    \item A CNN based sequential channel and CFO estimation is proposed and the benefit of the novel model investigated in terms of mean square error (MSE) and bit error rate (BER) for a wide signal-to-noise ratio (SNR) range along with common estimation techniques.
    \item  Emphasizing the capacity needs of large-scale networks, the power domain non-orthogonal multiple access (NOMA) method is tested on the proposed setup and the performance of the  CNN powered NOMA method was compared with orthogonal multiple access (OMA) systems. 

\end{enumerate}

\subsection{Related Literature}
 A recent study \cite{csenyuva2021harmonic} highlights the necessity of a more robust and versatile synchronization for LEO constellation and proposes to use the rotational invariance technique. A study \cite{le2021deep} proposes a CNN based channel estimation in tapped delay line (TDL) channel model and it shows the performance increment in terms of various estimation techniques. Another study \cite{faghani2020recurrent} states that long short-term memory of recurrent neural network (LSTM-RNN) structure has potential use for multiple input multiple output (MIMO) sparse channel estimation problems. Regarding synchronization, \cite{app10207267} finds packet arrival time using CNN architecture to drop false detection rates. An RNN approach to both CFO estimation and packet detection is shown and tested with software defined radios (SDR) in \cite{ninkovic2020deep}. Yet, a CNN-based sequential CFO and channel estimation study has not been conducted before. 
 \subsection{Organization of this paper}
 The outline of the study is as follows: In Section \rom{2}, channel and CFO effect on orthogonal  frequency-division multiplexing (OFDM) signal is analyzed and improved by CNN algorithm. In Section \rom{3}, the advantages and applicability of the NOMA technique over the current HAPS-LEO system are mentioned. In Section \rom{4}, CNN aided NOMA HAPS-LEO system is compared with OMA. As a conclusion, the inference of the authors' regarding the subject are shared in Section \rom{5}.
 
\section{Estimation Analysis of a Receiver}

\subsection{CFO Model and Estimation }


There is a substantial amount of CFO among space units because of the velocity of free falling nodes, causing Doppler shift. The speeds of the satellites vary depending on their orbital positions, so does the CFO. In addition to these reasons that exacerbate the shift, effects arise from the hardware difference of the receiver and transmitter and the time reference mismatch. Hence, it is necessary to examine an independent CFO model with the randomness on this system. 

Orthogonal frequency-division multiplexing (OFDM) is a digital transmission technique for encoding data on several subcarrier frequencies to achieve a system less susceptible to interference while providing more efficient data bandwidth. It is frequently used in cellular networks, mobile broadband standards and the next generation wireless LAN applications. 

Using OFDM waveform for a single input single output system case, $\kappa$ is the total number of subcarriers, $k$ denotes the subcarrier index and $n$ is the sample index of $N$-point IFFT taken signal. Index of OFDM symbol is denoted with $l$ whereas the total number of the packet is $\mathcal{L}$. 
Denoting $\{x_l[n]\}^{N-1}_{n=0}$ as transmitting samples and $y_l[n]$ is the received OFDM symbol, the effect of normalized CFO with subcarrier interval ($|\varepsilon|\leq0.5$) in the time domain can be formulated as
\begin{equation}
    y_l[n]= x_l[n]e^{j2\pi\varepsilon/N},
\end{equation}


For simplicity, exposing the discrete Fourier transformed (DFT) of an OFDM  signal $Y_l[k]$ to the channel can be done easily in the frequency domain where $H$ is the channel frequency response (CFR) of the received signal and $\mathcal{F}$ is DFT operator, $y_l[n]$ = $\mathcal{F}^{-1}\{Y_l[k]\}$ can be obtained as following
\begin{equation}
 y_l[n] = \frac{1}{N} \sum_{k=0}^{N-1} H_l[k]X_l[k]e^{j2\pi(k+\varepsilon)n/N}+ z_l(n).
\end{equation}
The CFO effect on the OFDM symbol is clearly signified at the receiver in the time domain. It is possible to examine this frequency change and residual effects in the frequency domain expressed as
\begin{equation} \label{eq4}
\begin{split}
& \mathcal{F}\{y_l[n]\} = \sum_{n=0}^{N-1}\frac{1}{N} \sum_{m=0}^{N-1} H_l[m]X_l[m]e^{j2\pi (m+\varepsilon)n/N} + \\
& \sum_{n=0}^{N-1}z_l[n]e^{-j2\pi kn/N},
\end{split}
\end{equation} 
The closed expression of  (\ref{eq4}) results in
\begin{equation}
   Y_l[k]=\frac{\sin(\pi \varepsilon)}{N\sin(\pi \varepsilon/N)}e^{j\pi \varepsilon(N-1)/N} H_l[k] X_l[k] + I_l[k] + Z_l[k].
\end{equation}
While the first element of  $Y_l[k]$ shows the weakening in power, the second element shows the cyclic shift of the received data. In the meantime, $I_l[k]$ is the element of the residual shift in subcarriers leading to intercarrier interference, and $Z_l[k]$ is the additive white gaussian noise with i.i.d. $\mathcal{CN}(0,N_0)$ elements.

In this study, 
instead of use of cyclic prefix or pilot tones, it was found more appropriate to use the  $||\mathcal{P}||$ length preamble segment, with the same logic of primary synchronization signal (PSS) and Shcmidl \& Cox \cite{schmidl1997robust} algorithm. As in this case, $\varepsilon_l$ can be defined as
\begin{equation}
    \varepsilon_l = \hat{\varepsilon}_l + \xi_l,
\end{equation}
where $\hat{\varepsilon}$ normalized estimation of CFO and $\xi$ is the normalized residual CFO. 
$\mathcal{P}$$_1[k]$ and $\mathcal{P}$$_2[k]$ are the two repetitive identical sequences of $n$-th OFDM symbols to find the $\hat{\varepsilon}$ $\in$ $\mathbb{R}$$^{1\times1}$ as 
 \begin{equation}
      \frac{\mathcal{P}_2[k]}{\mathcal{P}_1[k]} = e^{j2\pi \hat{\varepsilon}},
 \end{equation}
 and obtaining normalized angle by
\begin{equation} \label{eq:eps}
    \hat{\varepsilon}_l = \frac{1}{2\pi}\arctan \Bigg\{\frac{\sum_{k=0}^{N-1} \Im [{\mathcal{P}_1^*[k]}{\mathcal{P}_2[k]}]}{\sum_{k=0}^{N-1} \Re [{\mathcal{P}_1^*[k]}{\mathcal{P}_2[k]}]}\Bigg\}.
\end{equation}

\subsection{Channel Model and Estimation}
In a practical manner, the least-squares (LS) estimation technique is a highly effective and common method in the OMA case. The channel model of the HAPS-LEO network containing aerial communication link is considered as Rice fading channel due to LOS propagation \cite{stuber1996principles}. Hence, throughout the study,  this channel  model has been used  to illustrate HAPS-LEO channel condition.

Considering a usual OMA scenario, pilot tones $\textbf{X}$ = diag($\nu$): $\nu$ $\in$ $\mathbb{R}$ $^N$, $\textbf{H}$ is the channel coefficients and $\textbf{Y}$ = [$Y[1]$ $Y[2]$ $\ldots$ $Y[D]$]$^T$ is the received signal vector in FD, received data can be written as
\begin{equation}
\textbf{Y} = \textbf{X} \textbf{H} + \textbf{Z},
\end{equation}
\textbf{Z} is the noise for each subcarrier belonging to unique users. In a case where the pilot tone ratio is $f_p$, the result of the channel coefficients estimated to minimize the loss function through $N/f_p$ length $\textbf{H}$ can be found 
\begin{equation} \label{eq:10}
\frac{\partial ||\textbf{Y}-\textbf{X}\textbf{H'}||^2}{\partial \textbf{H'}}=0,
\end{equation}
which concludes with the solution of LS estimator as
\begin{equation} \label{eq:ls}
    \textbf{H'}_{\textbf{LS}} = (\textbf{X}^H\textbf{X})^{-1} \textbf{X}^H\textbf{Y} = \textbf{X}^{-1} \textbf{Y}.
\end{equation}
By using $\textbf{H'}_{\textbf{LS}}$, the received signal is equalized with the zero forcing method, and data reconstruction processes are completed. In addition, the minimum mean square error method, which performs channel estimation by exploiting the SNR value, is not very practical due to its high complexity.

\subsection{CNN Based Channel and CFO Estimation Model }

\begin{figure}[b]
    \centering
    \includegraphics[width=0.5\textwidth]{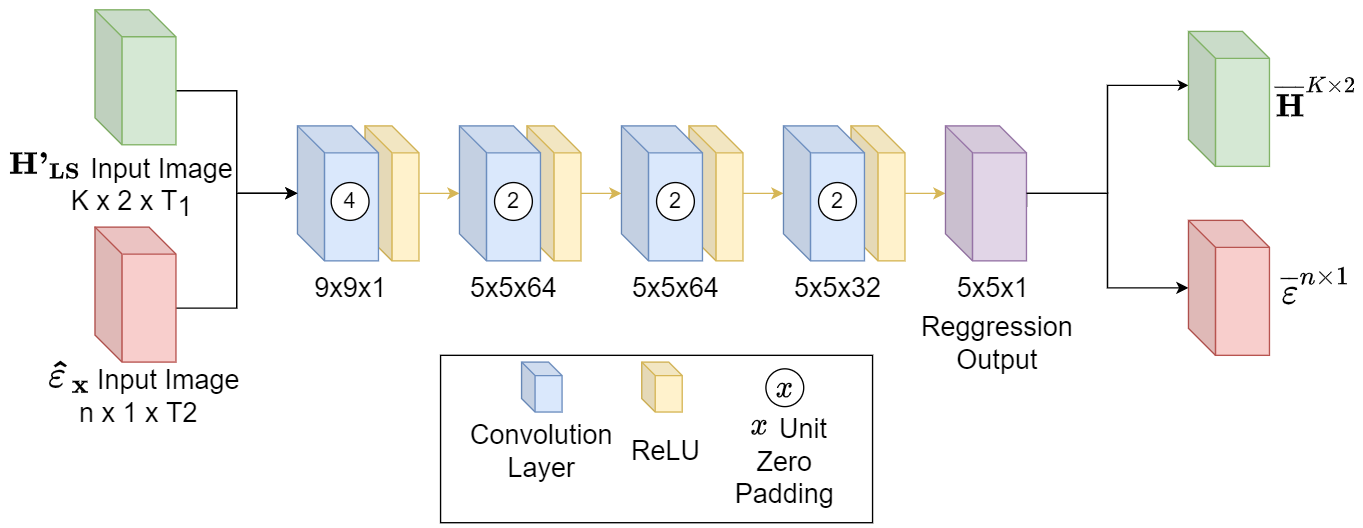}
    \caption{The proposed CNN Diagrams}
    \label{cnndiag}
\end{figure}



Signal denoising is the motivation of neural network usage for estimation purposes. In order to take full advantage of the power of neural networks and to keep complexity to a minimum, the use of CNN has been considered. 2D CNN structure is not only a powerful algorithm to capture spatial and temporal dependencies, but also has a light computation capability without losing high level features thanks to convolution layers. Layers including $p_q\times p_q$ convolution kernels take input with $p_{q-1}\times p_{q-1}$ size. Every convolution layer results with:
\begin{equation}
C_e = \textrm{Conv}(A_p,w_p) + b_p, 
\end{equation}
where Conv() is the convolution operator, $A$ is the input layer, $w$ is the weight and $b$ is the bias. After each convolution layer, the ReLU is used to regulate the outputs by setting negative values $``0"$
\begin{equation}
    R(z)=\max\{0,z\}.
\end{equation}
Since each channel coefficient is complex, real and imaginary parts gets involved in the training and evaluation procedure separately as input of the training sets are
\begin{gather} 
\textbf{X}^{\textbf{CNN}}_\textbf{H}=\{\Re\{\textbf{H'}_{\textbf{x}} \}, \Im\{\textbf{H'}_{\textbf{x}}\}\}, \\
\textbf{X}^{\textbf{CNN}}_\mathbf{\varepsilon}=\{\mathbf{\hat{\varepsilon}}_x\},
\end{gather}
and similarly the output
\begin{gather}
\textbf{Y}^{\textbf{CNN}}_\textbf{H} = \{\Re\{\overline{\textbf{H}}_{\textbf{CNN}} \} , \Im\{\overline{\textbf{H}}_{\textbf{CNN}}\}\}, \\
\textbf{Y}^{\textbf{CNN}}_\mathbf{\varepsilon}=\{\mathbf{\overline{\varepsilon}}_y\}.
\end{gather}
Approximation process with training to obtain \textbf{$\Theta$} that reach output vector as close as possible in terms of Euclidian distance metric as
\begin{equation}
    f (\Theta,\textbf{X}^{\textbf{CNN}},\textbf{Y}^{\textbf{CNN}}) \\
    = \frac{1}{T}\sum_{t=1}^T ||y_t^{CNN} - \hat{f}(x_t^{CNN} ; \Theta)||^2,
\end{equation}
where $\hat{f}$ is the generalization function, estimation of perfect estimator $f$, $\textbf{X}^{\textbf{CNN}}$ = [$x_1$ $x_2$ ... $x_t$] and $\textbf{Y}^{\textbf{CNN}}$ = [$y_1$ $y_2$ ... $y_t$].

Independency of $\hat{\varepsilon}_x$ and  ${\textbf{H'}_{\textbf{x}}}$
lets CNN estimators lead the solution by updating one parameter at a time. For these two parameters, the network minimizes the $\textbf{$\mathscr{F}$($\hat{\varepsilon}  ,  \breve{H}$)}$ operation by solving variable optimization problem for each vector of variables where $\breve{H}$ is $\overline{\varepsilon}$ deduced estimated channel that $\overline{\varepsilon}$=$\hat{\varepsilon}^{opt}$ and finalizing process with $\overline{H}$=$\widehat{H}^{opt}$.


As in training case, two parameters estimation starts with $\hat{\varepsilon}_x$ using its ground truth as prior knowledge $\varepsilon$ and completes with finding $\overline{H}$ $\in$ ${\overline{\textbf{H}}_{\textbf{CNN}}}$ by obtaining each loss parameters \textbf{$\Theta$} as following
\begin{equation}
    \widehat{\textbf{$\Theta$}}_\varepsilon = \operatorname*{argmin}_{{\hat{\varepsilon}\in{\Bbb R}}} f(\textbf{$\Theta$}_\varepsilon,\hat{\textbf{$\varepsilon$}}_x,\varepsilon),
\end{equation}
and
\begin{equation}
    \widehat{\textbf{$\Theta$}}_H = \operatorname*{argmin}_{{\widehat{H}\in{\Bbb C}}} f(\textbf{$\Theta$}_H,\widehat{H}_x,H),
\end{equation}
Subsequently, the evaluation section makes use of $\widehat{\textbf{$\Theta$}}_{\varepsilon,H}$ by using $\hat{\varepsilon}$ and $\breve{H}$ with $\overline{\varepsilon}^{n}_\delta$ =  $f(\hat{\varepsilon}$ ; $\widehat{\textbf{$\Theta$}}_\varepsilon)$ and $\overline{H}^{\kappa}$ = $f(\breve{H}$ ; $\widehat{\textbf{$\Theta$}}_H)$. CFO-CNN and channel estimation using CNN (CE-CNN) architectures with their kernel sizes can be seen from Figure \ref{cnndiag}. 

\subsubsection{Optimization of Gradient Descent}

Although SGD computation performs just as regular gradient descent algorithm, it behaves poorly for high learning rate. Adaptive moment estimation (Adam) optimizer \cite{kingma2014adam} uses the first ($m$) and second ($\nu$) moment of gradient to adapt learning rate for each $t$ iteration
\begin{gather}
m_{t+1} = \beta_1 m_{t} + (1-\beta_1)\frac{\partial  \widehat{\textbf{$\Theta$}}}{\partial w_t}, \\
\nu_{t+1}= \beta_2 \nu_t + (1-\beta_2) (\frac{\partial  \widehat{\textbf{$\Theta$}}}{\partial w_t})^2,
\end{gather}
where moving average and exponential moving average are controlled by the gradient decay factors $\beta_1$ and $\beta_2$. Since initial points $m_0$ and $v_0$ are ``0'', instant zero biases show up. Therefore, a bias controller takes place
\begin{gather}
\hat{m}_t=\frac{m_t}{1-\beta_1^t} ,\\
\hat{\nu}_t=\frac{\nu_t}{1-\beta_2^t} ,
\end{gather}
With the found momentums, Adam weight update rule is applied instead of SGD
\begin{equation}
w_{t+1} = w_t - \hat{m}_t \frac{\eta}{\sqrt{\hat{\nu}}+\epsilon},
\end{equation}
where $\epsilon$ is the constant that keeps the bias from going to ``0'' and $\eta$ is the step size.

\section{NOMA for HAPS-LEO Links}

With the expectation of incoming mega constellation networks, a quite serious factor that will limit the connectivity performance of HAPS-LEO systems is congestion.
Number of mega constellation nodes that share work or fully satellite network dependent overcrowded systems may require more than HAPS aided network in an emergency event. Considering limited queueing models, the performance of HAPS-LEO networks that are utilized for use cases with high connectivity density can be restrained by data rate, in parallel with this, low channel capacity might become a threat. In the solution of this issue, the use of NOMA is an eye-catching method that will increase the efficiency of the spectrum by power allocation according to the channel structure of the users and increase the throughput.
By using  the  same frequency and time domain by providing spectrum efficiency, power domain NOMA, is a strong candidate for multiple access methods for large coverage network models that serve different  purposes. Furthermore, the biggest disadvantage of NOMA is eliminated by CNN estimation technique for this system by the reduction of user detection imperfection depending on the channel state information (CSI) quality. It has been noticed that HAPS-LEO network capacity and data throughput can be increased opportunistly and resulting with $\sum_{i=1}^{M}\textrm{log}_2(1+  \text{SNR}_i)$ total sum rate that makes a HAPS-LEO network immensely both power and spectral efficient where $M$ is the total number of users and $i$ is the users' index.


Strong users with their own power coefficient knowledge need a multiuser detection algorithm to extract their data. For this study, this algorithm was chosen as the conventional successive interference cancellation (SIC). Therefore, the strong user's quality of SIC operation is directly related to the CSI knowledge and synchronization quality. The fundamental order of NOMA steps is based on the authors' previous experimental study that accommodates 4 NOMA users \cite{durmaz2020four} and the NOMA parameters are selected in accordance with this deployment.

\subsection{NOMA Downlink}

NOMA is a contemporary method that creates band and energy efficiency, establishes user fairing by power allocation. While low power is assigned to the user (strong user) whose channel performance is above the appropriate criteria, high power is assigned to the user with poor channel condition (weak user). After modulation, the merged $M$ users in the transmitter creates a single superimposed signal as follows

\begin{equation}
    y_t = \sum_{i=1}^{M} \sqrt{a_i P_t}x_i,
\end{equation}
where $x_i$ is the modulated data of the users, $a_i$ is the power coefficient of $i$'th user and $P_t$ is the total power at the source that will be distributed to each user afterward. Defining the channel impulse response of any user as $h$, received signal at any user $\Tilde{x}$ is
\begin{equation}
    \Tilde{x} = y_t + z = \sum_{i=1}^{M} \sqrt{a_i P_t}x_i h_i +z_i,
\end{equation}
where $i$ is the NOMA user index and $z$ $\in$  $\mathcal{CN}(0,N_0)$. 

The received $\Tilde{x}$ gets equalized with $\textbf{H'}_{\textbf{LS}}$, obtaining $\overline{x}_i$ which is the received signal by the $i$-th user in the same manner. The weak user passes through the decoding stage without being exposed to any SIC, meaning $x_1$=$\overline{x}_1$. Yet, notating $U_2$ detected first user as $\hat{x}_2$, strong users extract their data from the received signal as following
\begin{equation}
x_2 = \frac{\overline{x}_2-\hat{x}_1\sqrt{\alpha_1}}{\sqrt{(1-\alpha_1-\alpha_3)}},
\end{equation}
and $U_3$ cancels the remaining users interference with one step forward 
\begin{equation}
x_3=\frac{\overline{x}_3-\hat{x}_1\sqrt{\alpha_1}-\hat{x}_2\sqrt{\alpha_2}}{\sqrt{(1-\alpha_1-\alpha_2) }}.
\end{equation}
Note that, as the nature of NOMA, the more SIC layers results with higher computational complexity in users.  

\subsection{NOMA Uplink}

Uplink communication occurs by combining the signals of users with unique channels in the BS and extracting the data of each user using SIC. In case of user signal output is $\sqrt{a_i P_t}x_i$, the received signal at the BS can be modeled as
\begin{equation}
     x_t  = \sum_{i=1}^{M} \sqrt{a_i P_t}x_i h_i +z_i,
\end{equation}
where $h_i$ $\in$ $\mathbb{C}$$^{N\times 1}$ is the $i$-th user's channel impulse response. On the contrary of DL, UL NOMA signal includes $M$ amount of channel with possibly different distributions. 

One of the important points is that the strong user (for our case $U_3$) cannot go beyond being noise to the user with weak and high power coefficients. This means that while the channel estimation of the already weak user is vital, the strong user whose signal strength is reduced by the power distribution should also make an effective estimation. 

As in the UL-NOMA case with a single CFO term, Eq. (\ref{eq:eps}) and (\ref{eq:ls}) are valid as $\hat{\varepsilon}_{i,l}$ = $\hat{\varepsilon}_l$. Regarding the UL-NOMA channel estimation case, gathered data from $M$ users channel considered as a single artificial channel $\dddot{\textbf{H}}$ at the BS and the estimation can be found just as a single user OMA user by LS estimator as in Eq. (\ref{eq:10}). The BS utilizes same channel for all users as $\overline{x}_1$=$\overline{x}_2$=$\overline{x}_3$. Hence, every user posseses different channels, the SIC error is expected to be significant. 
\section{Numerical Results}
\subsection{Performance Analysis for CFO and Channel  Estimation}

\begin{table}[]
	\centering
	\caption{Training Parameters}
	\label{table:tra}
	\resizebox{0.48\textwidth}{!}{%
		\begin{tabular}{|@{}l|c||c@{}l|c}
			\toprule
			\multicolumn{3}{c|}{Parameters} & \hspace{3mm} {CE-CNN} & {CFO-CNN} \\  \hline 
			\multicolumn{3}{l|}{Training function} &  \hspace{1.5mm} SGD(Adam)  & SGD(Adam) \\ \hline
			\multicolumn{3}{l|}{Maximum number of epoches}    & \hspace{7mm} 10  & 60  \\ \hline
			\multicolumn{3}{l|}{Mini-batch size}   & \hspace{7mm} 32 & 8\\ \hline
			\multicolumn{3}{l|}{Gradient Decay Factors ($\beta_1$, $\beta_2$)}   & \hspace{2mm} 0.9, 0.999 & 0.9, 0.999\\ \hline

			\multicolumn{3}{l|}{Learning Rate ($\eta$)}   &\hspace{4mm}  0.0005 & 0.0005\\ \hline
			\multicolumn{3}{l|}{Bias constant ($\epsilon$)}   &\hspace{5mm}  $10^{-8}$ & $10^{-8}$\\ \hline
			\multicolumn{3}{l|}{Validation from $\textbf{X}^{\textbf{CNN}}_\textbf{H}$}    & \hspace{6mm} 15\%  & 15\%  \\ \hline  
			\multicolumn{3}{l|}{Number of training samples ($T_{i}$)}   & \hspace{5mm} 10000 & 10000\\ 
			\bottomrule
				
		\end{tabular}%
}
\end{table}

\begin{figure}[b] \label{fig:mse}
    \centering
    \includegraphics[width=0.45\textwidth]{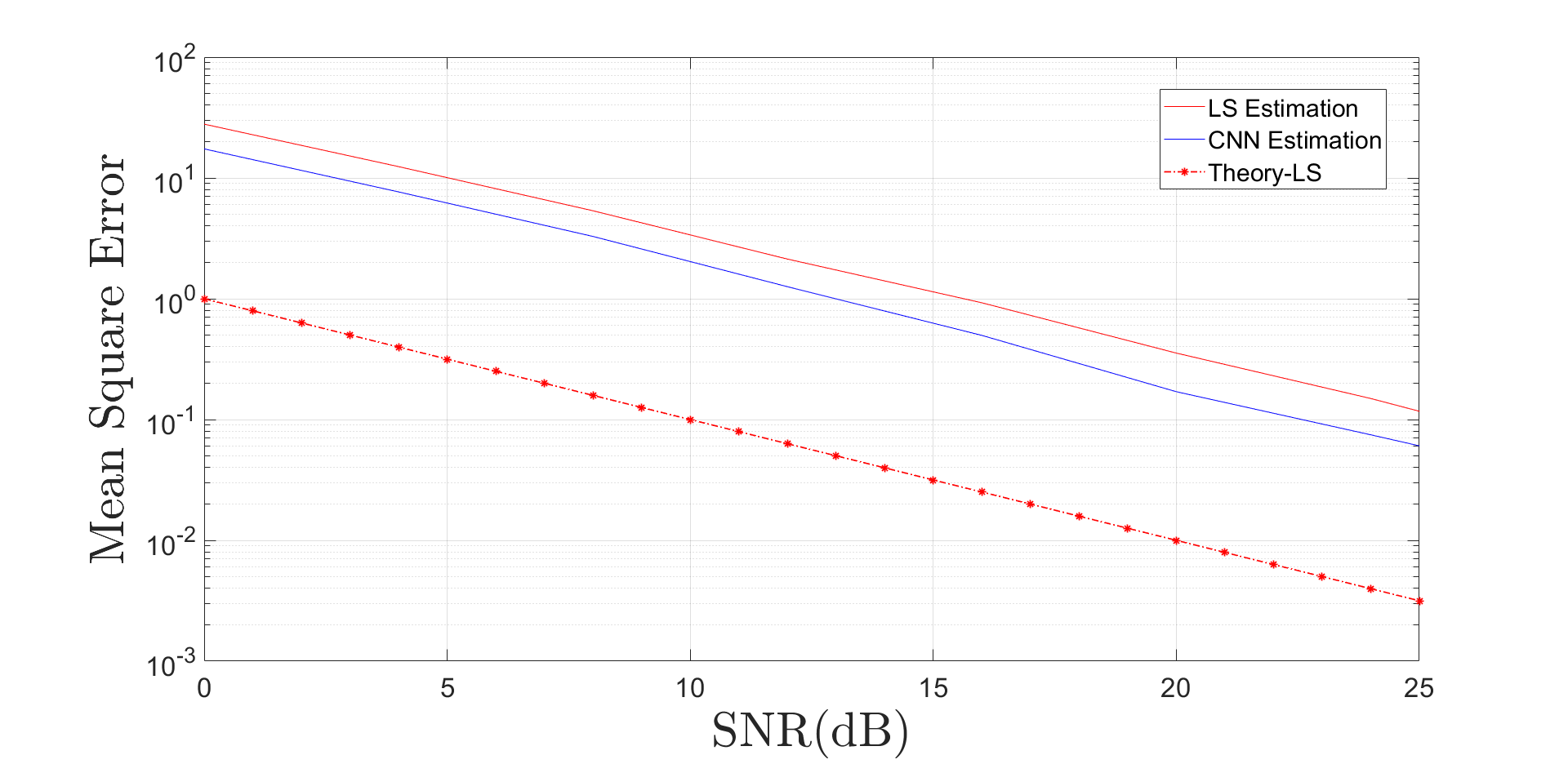}
    \caption{Comparison of simulations and theoretic MSE.}
    \label{MSESIM}
\end{figure}

Two independent CNN are used to regression prediction with stochastic gradient descent (SGD) approach. CE-CNN and CFO-CNN input data is produced separately. The receiver, which is exposed to HAPS-LEO defined channel and $\varepsilon_x$, estimates each OFDM symbol channel ${\textbf{H'}_{\textbf{x}}}$ for using the LS estimator for the CE-CNN case. Moreover, normally distributed pseudorandom input bits are considered as data and the OFDM subcarriers are generated with IDFT of modulated input data. As can be seen, both estimators use a common kernel structure, which is simple enough in both training and model output.

Both designs are trained with premade synthetic data with uniformly random SNR values between 5 to 15 dB. As in the CE-CNN case, 10,000 $\textbf{H'}_{\textbf{x}}$ $\in$ $\mathbb{C}$$^{128\times2}$ were taken as input with assigning 15\% of it as validation data. The input passed through 64 units of 9$\times$9 Conv filters using ReLU with 1 unit of zero padding, resulting in 128$\times$2$\times$64 in the first convolution layer output. 
As in the output layer, the single regression layer computes the mean squared error right after the latest ReLU activation to end the training session. 



Similar to CE-CNN, CFO-CNN case input of 10,000 $\hat{\varepsilon}_x$ $\in$ $\mathbb{R}$ $^{1100\times1}$ is generated with by obtaining $\hat{\varepsilon}_x$ in a trial simulation under Rician fading and ${\varepsilon}$ CFO. Unlike CE-CNN, since the dynamic state of $\varepsilon$, the epoch iteration number is increased for the CFO-CNN process. In order to simulate a time varying space environment, ${\varepsilon}$ $\in$ $\sum_{i=1}^{U}\phi_i\times\mathcal{N}(\mu_i,\Sigma_i)$ is created as time varying with a gaussian mixture distribution with random $\phi$ weights. All parameters for CNN structure and link design is shown in Table \ref{table:tra} and \ref{tab:table1}.
Rician fading channel is considered in all cases with $K$=10 factor. Frequency selective $L$=3 taps channel structure was used instead of flat fading which is another challenge that is thought to be important in benefiting from CNN. Unit power pilot tones are interleaved into the OFDM symbols regularly with $p_f$=1:8. Preambles of alternating repetitive sequence lengths are ||$\mathcal{P}_1$|| = ||$\mathcal{P}_2$|| = 160.

\begin{figure}[] \label{fig:res}
    \centering
    \includegraphics[width=0.45\textwidth]{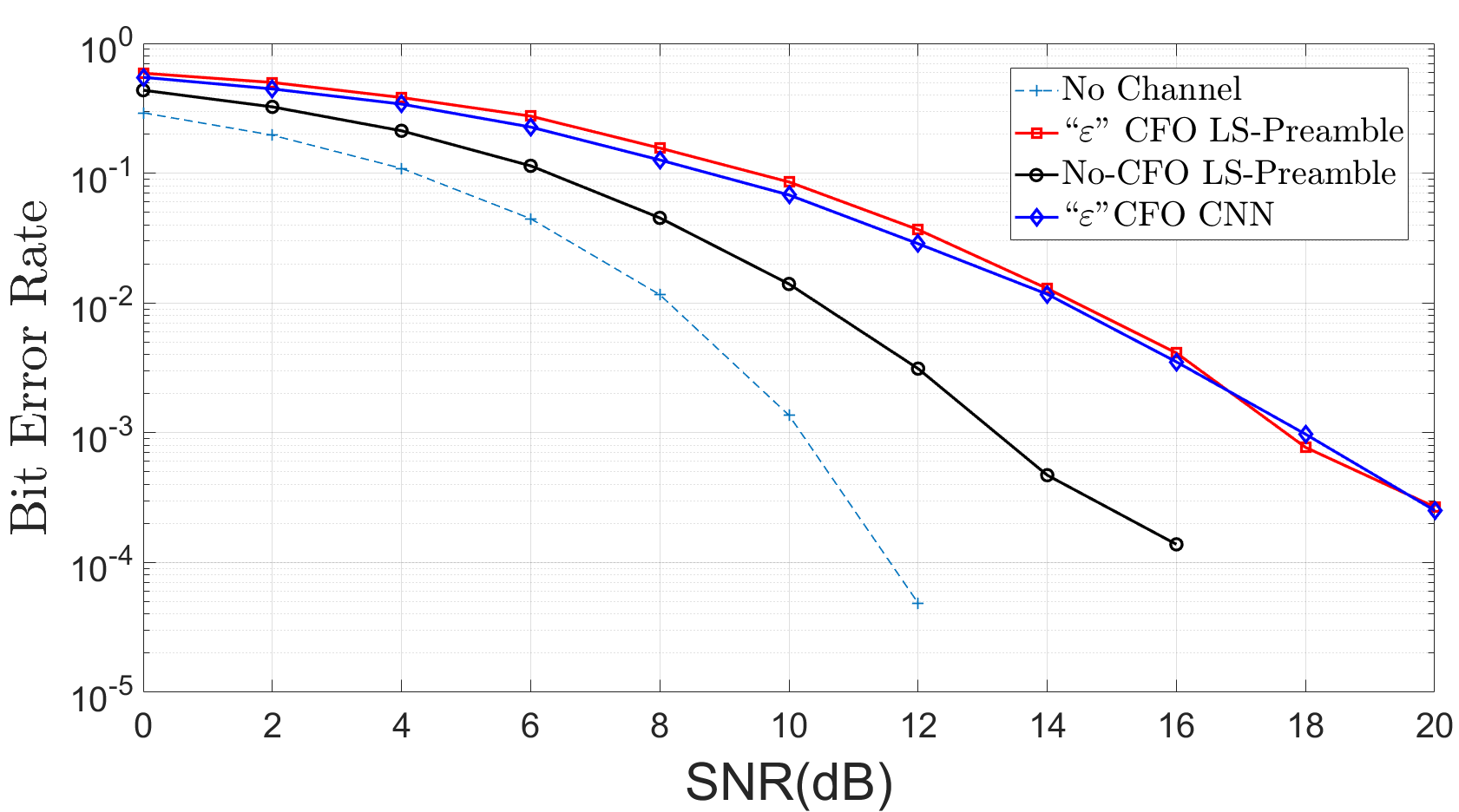}
    \caption{The effect of the channel on the $\varepsilon$ and the comparison of CNN with the Preamble method in terms of BER.}
    \label{res}
\end{figure}

\begin{figure}[b] \label{fig:3deps}
    \centering
    \includegraphics[width=0.45\textwidth]{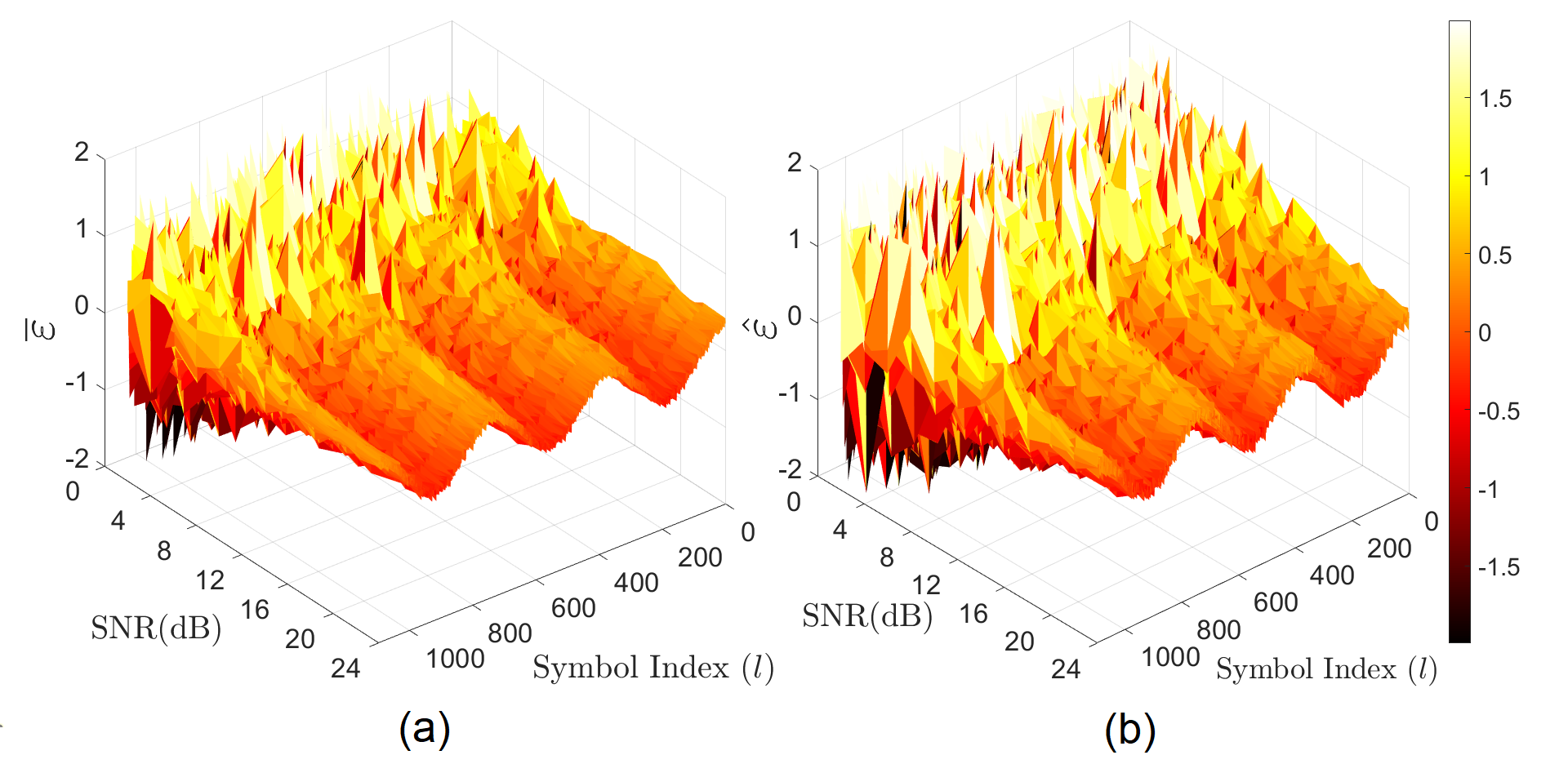}
    \caption{CNN vs Preamble estimation for for 0-24 dB SNR.}
    \label{3deps}
\end{figure}

Figure \ref{MSESIM} shows that MSE for CFO and channel exposed environment CNN supported signal gets to equalize with $\overline{H}$ where $\sigma^2_{\overline{H}}$ $<$ $\sigma^2_{\widehat{H}}$ for a wide SNR range. A noteworthy observation, in the fast fading points of the channel, CNN can adapt the interpolated intermediate channel coefficient values better than the LS method. 
The vitality of the aforementioned $\xi$ effect can be seen in the practical scenario of Figure \ref{res} involving it, the SNR loss caused by $\xi$ itself and the effect of the channel estimation error it causes. Reflections of accumulated CNN estimation errors to bit errors are lower than the LS-Preamble case shown in Figure \ref{res}.

The density of CFO magnitudes, which are thought to exceed half of the subcarrier interval, and packet losses are seen in 3-dimensional Figure \ref{3deps} where \ref{3deps}(a) represents the CNN aided CFO estimations for each OFDM symbols and \ref{3deps}(b) shows the only LS/Preamble $\hat{\varepsilon}$ results. It can be seen that packet losses in the low SNR region are reduced by using CNN.
\begin{figure}[t]
    \centering
    \includegraphics[width=0.45\textwidth]{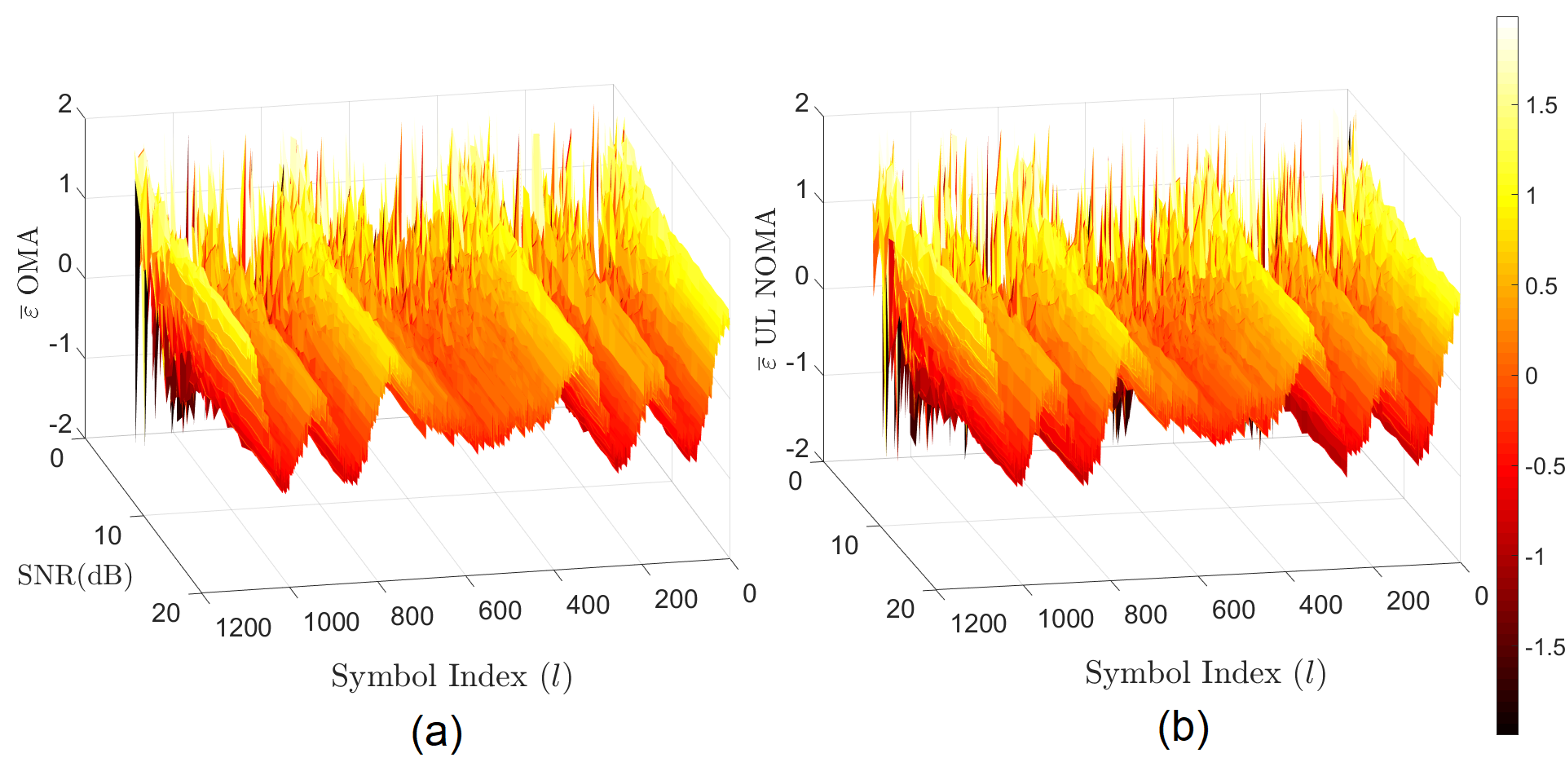}
    \caption{OMA vs NOMA CNN Estimation Accuracy for 0-20 dB SNR range.}
    \label{MSEUL}
\end{figure}
\vspace{-0.3cm}
\subsection{NOMA-OMA Comparison}

\begin{table}[b]  
\begin{center}  

\caption{System Parameters}  
\label{tab:table1}  
\begin{tabular}{p{1 cm}|c|r}  
\toprule 
  
\textbf{Design} & \textbf{Parameters} & \textbf{Values}\\  
\midrule 

  & NFFT ($N$) & 256\\\cline{2-3} 
  & \# of Subcarriers ($\kappa$)& 128\\\cline{2-3}  
  & CP Length  & 16\\\cline{2-3}
  & Modulation  & 4-QAM\\\cline{2-3} 
  & \# of Paths ($L$) & 3\\\cline{2-3} 
  & Preamble Length (||$\mathcal{P}$||)& 160 \\\cline{2-3}
  OFDM & Carrier Frequency & 1.2 GHz\\\cline{2-3} 
  & Pilot Ratio ($f_p$)& 1/8 \\\cline{2-3}
  & \# of Symbols ($\mathcal{L}$)  & {\tiny $10^5$}\\\cline{2-3} 
  & Channel Model & Rician Fading, $K$=10\\\cline{2-3} 
  & CFO Model  & Gaussian Mixture\\\cline{2-3} 
  & Noise Model & Additive White Gaussian\\   
  
\midrule 

& \# of NOMA users ($M$) & 3 \\\cline{2-3} 
NOMA & NOMA Coefficients ($\alpha_i$) & 0.761, 0.191, 0.048 \\ \cline{2-3}
& User Detection & \small{SIC}  \\

\bottomrule
\end{tabular}  
\end{center}  
\end{table}

UL-NOMA and DL-NOMA network built on the foregoing OMA scenario has been simulated for $M$=3 users. Fixed power coefficients are selected as \cite{durmaz2020four} states. MSE and BER results of NOMA tests using the same OMA simulation parameters were obtained for both CNN and non-CNN cases.  Figure \ref{MSEUL}, 
shows $\overline{\varepsilon}$ for 1100 sample test symbols. As a result of our observations, residual CFO ($\xi$) seems to be negligible for signals reached above 10 dB, while the amount of error dramatically increases for all cases below 10 dB.

\begin{figure}[t]
    \centering
    \includegraphics[width=0.45\textwidth]{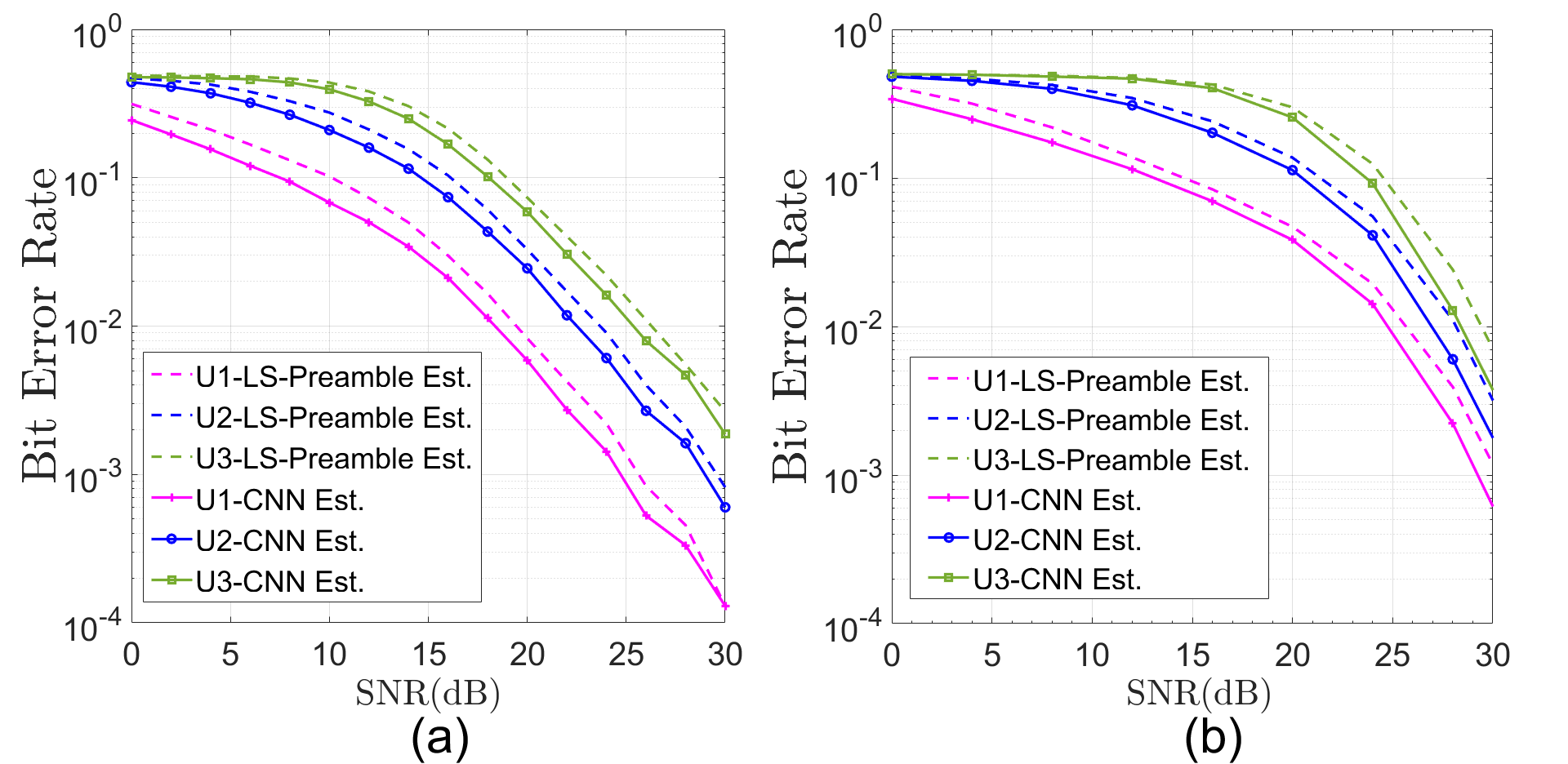}
    \caption{CNN-aided Downlink NOMA System.}
    \label{BERDL}
\end{figure}

Figure \ref{BERDL}(a) shows that a single CNN supported HAPS BS outperforms the non-CNN case where the reconstruction procedure utilizes only LS/Preamble algorithms. Figure \ref{BERDL}(b) illustrates the comparison of CNN used UL-NOMA and only LS/Preamble estimated UL-NOMA cases. To briefly compare with the DL-NOMA case, UL-NOMA data reconstruction is highly critical because 3 different users were exposed to independent $H_i$ and $\varepsilon_i$. Fortunately, the CNN structure, as in the case of DL-NOMA, lowered the BER by providing more accurate $\varepsilon$ and in all ranges from 0 to 30 dB SNR. NOMA parameters are summarized in Table \ref{tab:table1}.

\section{Conclusion}
The solutions that a communication network with HAPS-LEO infrastructure can provide to mobility and load problems are discussed and the importance of channel estimation and synchronization problems for both UL and DL are justified by different proofs. To overcome this issues, a CNN based solution is proposed. It has been observed that the CNN, indeed, minimizes estimation errors and improves both MSE and BER performances. The capacity problem required for wide coverage systems was tried to be solved by using the proven NOMA method and the installability of the 3 users UL/DL-NOMA HAPS-LEO system was proven with BER curves. Finally, the CNN-assisted performance of the NOMA model was examined in comparison with OMA, and it was shown that it achieved low BER for all users over a wide range of SNR's.

\vspace{0.5cm}
\bibliographystyle{IEEEtran}
\bibliography{refer}{}

\begin{thebibliography}{10}
\providecommand{\url}[1]{#1}
\csname url@samestyle\endcsname
\providecommand{\newblock}{\relax}
\providecommand{\bibinfo}[2]{#2}
\providecommand{\BIBentrySTDinterwordspacing}{\spaceskip=0pt\relax}
\providecommand{\BIBentryALTinterwordstretchfactor}{4}
\providecommand{\BIBentryALTinterwordspacing}{\spaceskip=\fontdimen2\font plus
\BIBentryALTinterwordstretchfactor\fontdimen3\font minus
  \fontdimen4\font\relax}
\providecommand{\BIBforeignlanguage}[2]{{%
\expandafter\ifx\csname l@#1\endcsname\relax
\typeout{** WARNING: IEEEtran.bst: No hyphenation pattern has been}%
\typeout{** loaded for the language `#1'. Using the pattern for}%
\typeout{** the default language instead.}%
\else
\language=\csname l@#1\endcsname
\fi
#2}}
\providecommand{\BIBdecl}{\relax}
\BIBdecl

\bibitem{daehnick2020large}
C.~Daehnick \emph{et~al.}, ``Large {LEO} satellite constellations: Will it be
  different this time,'' \emph{McKinsey \& Company, https://www. mckinsey.
  com}, vol.~4, 2020.

\bibitem{CHEN2019376}
Q.~Chen \emph{et~al.}, ``A distributed congestion avoidance routing algorithm
  in mega-constellation network with multi-gateway,'' \emph{Acta Astronautica},
  vol. 162, pp. 376--387, 2019.

\bibitem{kurt2021vision}
G.~K. Kurt \emph{et~al.}, ``A vision and framework for the high altitude
  platform station (haps) networks of the future,'' \emph{IEEE Communications
  Surveys \& Tutorials}, vol.~23, no.~2, pp. 729--779, 2021.

\bibitem{wang2019effect}
X.~Wang \emph{et~al.}, ``The effect of {HAPS} unstable movement on handover
  performance,'' in \emph{IEEE Wireless and Optical Comm. Conf. (WOCC)}, 2019,
  pp. 1--5.

\bibitem{alzenad2018fso}
M.~Alzenad \emph{et~al.}, ``{FSO}-based vertical backhaul/fronthaul framework
  for {5G}+ wireless networks,'' \emph{IEEE Communications Magazine}, vol.~56,
  no.~1, pp. 218--224, 2018.

\bibitem{hsieh2020uav}
F.~Hsieh \emph{et~al.}, ``{UAV}-based multi-cell {HAPS} communication: System
  design and performance evaluation,'' in \emph{IEEE GLOBECOM}, 2020, pp. 1--6.

\bibitem{3gpp.36.331}
3GPP, ``{Unmanned aerial systems over 5G},'' 3rd Generation Partnership Project
  (3GPP), Technical Specification (TS), 2019, version 14.2.2.

\bibitem{itu371.1996}
ITU-T, ``Propagation data and prediction methods for systems using high
  altitude platform stations and other elevated stations in the stratosphere at
  frequencies greater than about 0.7 {GHz},'' Int. Telecommunication Union,
  Recommendation P.1409-2, 2021.

\bibitem{lecun1999object}
Y.~LeCun \emph{et~al.}, ``Object recognition with gradient-based learning,'' in
  \emph{Shape, contour and grouping in computer vision}.\hskip 1em plus 0.5em
  minus 0.4em\relax Springer, 1999, pp. 319--345.

\bibitem{csenyuva2021harmonic}
R.~V. {\c{S}}enyuva and G.~K. Kurt, ``Harmonic retrieval of {CFO} and frame
  misalignment for {OFDM}-based inter-satellite links,'' in \emph{IEEE Int.
  Symp. on Wireless Comm. Systems (ISWCS)}, 2021, pp. 1--5.

\bibitem{le2021deep}
A.~Le~Ha \emph{et~al.}, ``Deep learning-aided {5G} channel estimation,'' in
  \emph{Int. Conf. on Ubiquitous Info. Man. and Comm. (IMCOM)}, 2021, pp. 1--7.

\bibitem{faghani2020recurrent}
T.~Faghani \emph{et~al.}, ``Recurrent neural network channel estimation using
  measured massive {MIMO} data,'' in \emph{IEEE PIMRC}, 2020, pp. 1--5.

\bibitem{app10207267}
E.-R. Jeong \emph{et~al.}, ``Convolutional neural network {(CNN)}-based frame
  synchronization method,'' \emph{Applied Sciences}, vol.~10, no.~20, 2020.

\bibitem{ninkovic2020deep}
V.~Ninkovic \emph{et~al.}, ``Deep learning based packet detection and carrier
  frequency offset estimation in {IEEE} 802.11 ah,'' \emph{arXiv preprint
  arXiv:2004.11716}, 2020.

\bibitem{schmidl1997robust}
T.~M. Schmidl and D.~C. Cox, ``Robust frequency and timing synchronization for
  {OFDM},'' \emph{IEEE Trans. on Communications}, vol.~45, no.~12, pp.
  1613--1621, 1997.

\bibitem{stuber1996principles}
G.~L. St{\"u}ber and G.~L. Steuber, \emph{Principles of Mobile
  Communication}.\hskip 1em plus 0.5em minus 0.4em\relax Springer, 1996,
  vol.~2.

\bibitem{kingma2014adam}
D.~P. Kingma and J.~Ba, ``Adam: A method for stochastic optimization,''
  \emph{arXiv preprint arXiv:1412.6980}, 2014.

\bibitem{durmaz2020four}
M.~A. Durmaz \emph{et~al.}, ``A four-user non-orthogonal multiple access system
  implementation in software defined radios,'' in \emph{IEEE Int. Black Sea
  Conf. on Comm. and Networking (BlackSeaCom)}, 2020.

\end{thebibliography}

\end{document}